\documentclass[11pt,a4paper]{article}

\usepackage{jheppub}
\usepackage{amsmath}
\usepackage{amssymb}
\usepackage{amsfonts}
\usepackage{graphicx}

\graphicspath{{./Figs_reduced/}} 

\newcommand{\bR}{\ensuremath{\mathbb{R}}}

\newcommand{\bZ}{\ensuremath{\mathbb{Z}}}

\newcommand{\scG}{\ensuremath{\mathcal{G}}}

\newcommand{\scI}{\ensuremath{\mathcal{I}}}

\newcommand{\scN}{\ensuremath{\mathcal{N}}}

\newcommand{\scT}{\ensuremath{\mathcal{T}}}

\newcommand{\bea}{\begin{equation}\begin{aligned}}
\newcommand{\eea}{\end{aligned}\end{equation}}
\newcommand{\beq}{\begin{eqnarray}}
\newcommand{\eeq}{\end{eqnarray}}

\newcommand{\Gr}{\ensuremath{\textrm{Gr}}}

\newcommand{\Fig}{Fig.~\!}

\DeclareMathOperator{\Tr}{Tr}

\title{
Network and Seiberg Duality
}

\author[\clubsuit]{Dan Xie}
\author[\spadesuit]{and Masahito Yamazaki}

\affiliation[\clubsuit]{Institute for Advanced Study, Princeton, NJ 08540, USA}
\affiliation[\spadesuit]{Princeton Center for Theoretical Science,  Princeton University, Princeton NJ 08544, USA}

\bigskip

\abstract{
We define and study a new class of 4d $\scN=1$ superconformal
quiver gauge theories associated with a planar bipartite network.
While UV description is not unique due to Seiberg duality, 
we can classify the IR fixed points of the theory by a permutation,
or equivalently a cell of the totally non-negative Grassmannian. The story 
is similar to a bipartite network on the torus classified by a Newton polygon.
We then generalize the network to a general bordered Riemann surface and
 define IR SCFT 
from the geometric data of a Riemann surface.
We also comment on IR R-charges 
and superconformal indices 
of our theories.
}

\emailAdd{dxie@ias.edu}
\emailAdd{masahito@princeton.edu}

\begin{document}
\maketitle

\section{Introduction}

The view on constructing and studying superconformal field theory (SCFT) 
has changed dramatically over the last decades.
Conventionally a SCFT is defined by specifying the matter content and the Lagrangian.
This method, however, 
has a number of limitations. First, it is in many cases hard
to compute physical observables in the strongly coupled region of the parameter space.
Second, there are many qualitative properties of the theory not present in the Lagrangian, 
such as hidden symmetries which 
cannot be identified from the Lagrangian.\footnote{One famous example is the dual superconformal symmetry of 
$\mathcal{N}=4$ SYM theory \cite{Drummond:2008vq},
and another is the 3d mirror symmetry \cite{Intriligator:1996ex}, 
where there is no Lagrangian description which manifests all the non-Abelian global symmetries
acting on the Higgs and Coulomb branches.}
Third, the Lagrangian description is far from unique and it could take completely 
different forms in different corners of the parameter space, and the equivalence of the different descriptions 
requires highly non-trivial duality properties.  
Moreover, there are many interesting strongly coupled field theory which one could not even write the Lagrangian in any corner, 
and even if you have the Lagrangian it is not really useful 
if it is complicated (e.g. huge gauge groups and many matters).

In the middle nineties, brane/geometric constructions are used extensively in the study of quantum field theories, 
and it has been discovered that highly non-trivial dualities of field theories have very nice geometric representation,
for example as a rearrangement of the 
brane configuration \cite{Hanany:1996ie}.

More recently, a fresh perspective on quantum field theory 
has been emerging.
 The first example
is the Gaiotto's construction on four dimensional $\mathcal{N}=2$ SCFTs \cite{Gaiotto:2009we}. 
Instead of specifying a Lagrangian, one specifies
a genus $g$ Riemann surface and  punctures with specification of the local data. 
 One immediate virtue is that all
the UV deformations are encoded as geometric parameters of the Riemann surface and the local data defining the punctures.
Moreover, the IR theory is solved by the moduli space of the Hitchin's
 equation defined on the surface \cite{Gaiotto:2008cd,Gaiotto:2009hg}.
What is remarkable is that highly non-trivial S-duality of the field theory is trivialized; it corresponds to different 
pants decompositions of the same Riemann surface.  Moreover, one only needs to know the S-duality behavior 
of the four punctured sphere. Once we understand this simple local piece, the whole S-duality behavior is understood
by gluing/gauging this basic building block.

The second example is the three-dimensional $\mathcal{N}=2$ SCFT
constructed from a 
hyperbolic 3-manifold \cite{Terashima:2011qi,Terashima:2011xe,Dimofte:2011jd,Dimofte:2011ju,Cecotti:2011iy,Dimofte:2011py}.
Here, the 3d gauge theory is specified by an ideal triangulation and the choice of the polarization,
and a gluing of tetrahedra is translated into a gauging of global symmetries in gauge theories.
Different ideal triangulations are related by a sequence of local moves (2-3 moves),
and each move corresponds to a 3d mirror symmetry.
Moreover, the IR behavior 
is determined by (quantization of) the moduli space of the flat connections defined on the 3-manifold.

The third example is the four-dimensional $\scN=1$ SCFT described by bipartite graphs on a two-dimensional torus (see \cite{Hanany:2005ve,Franco:2005rj,Franco:2005sm}).
The global data here is a toric Calabi-Yau 3-manifold determined from a convex polygon, and 
different bipartite graphs corresponding to the same geometry are related by a chain of Seiberg dualities 
\cite{Beasley:2001zp}, which are realized as square moves on the graph. 
The theories originates from the compactification of an NS5-brane intersecting with D5-branes 
(\cite{Imamura:2007dc,Yamazaki:2008bt}, see also \cite{Feng:2005gw}).

While all these developments start with global geometric objects and the duality frames 
are understood as the different decompositions of the same object, physically it is easier
to take a bottom-up approach; start with simple matter systems and gauge the flavor symmetries together to build 
complicated theories, and study the IR fixed point defined by the new theory.
However, there are several questions about this approach. 
First, can we define a simple geometric object associated with the field theory so
it is easy to study them? Second, the UV Lagrangian 
descriptions might be redundant, i.e. some of the gauge groups are decoupled in the 
IR. Can we find a minimal duality frame in which nothing is redundant?
Third, it is hard to judge whether two complicated theories are related by a sequence of 
dualities, it would be nice to get a global data independent of the duality frame.

The goal of this paper is to provide a concrete answer to these questions,
in a class of four dimensional $\mathcal{N}=1$ SCFTs.  
The geometric object is a planar bicolored network which is a generalization of the dimer model of the torus to the disc.
Instead of starting from some global data, we just start with a random bicolored network which physically
means we just randomly gauging the matter systems. There are several operations on the network which are perfectly 
mapped to the operations on the physical operations which do not change the IR fixed point.  Therefore the 
physical questions are mapped to the combinatorial question of the network \cite{PostnikovTotal}. 

The minimal duality frames are identified with the reduced network and there are easy combinatorial ways of 
judging whether or not a theory is minimal, and of finding the reduced
one. Similarly to the dimer story,
 different geometrical/combinatorial representations are related by 
Seiberg dualities \cite{Seiberg:1994pq}, which are trivialized by the so-called local square move on the planar bipartite network. 
The duality equivalence class of our SCFTs, when the bicolored graphs are reduced,
are classified by combinatorial data of a permutation (or equivalently a cell of the totally non-negative
Grassmannian $(\Gr_{k,n}(\bR))_{\ge 0}$ \cite{PostnikovTotal}), which are the invariant data independent of the duality frame).

We also extended our study to bordered Riemann surfaces (section \ref{sec.def1}).
Here we take a top-down approach; we start with the global data, i.e. a
Riemann surface, and 
identify the minimal duality frames. 
It is further shown that this class of theories could again
be described by bipartite networks, 
this time defined on a higher genus Riemann surface. 
Therefore all the theories in this paper are defined from the network and 
in each case we have identified the global data characterizing the IR
fixed point, as well as the minimal duality frames.

We will comment on 
the physical properties of our 
4d $\scN=1$ SCFTs (section \ref{sec.properties}),
including the IR R-charge and the 4d
superconformal index.

Towards the completion of this work we became aware of another work
by S.~Franco, which has some overlap with our work \cite{FrancoDraft}. We have coordinated submission of our papers.

\section{Planar Bicolored Network}\label{sec.def2}
In this section, we will define a quiver gauge theory from a reduced
planar bicolored network\footnote{In \cite{PostnikovTotal}, 
a network means a graph together with weights associated to the 
edges or faces of the graph. The weighs are essential in \cite{PostnikovTotal}
but will not be necessary for the definition of the theory in this paper, hence our terminology here
is that a network is synonymous with a graph.}.
The operations on 4d gauge theories keeping the IR fixed point, including the Seiberg duality, 
will be interpreted as combinatorial operations on the network. We will furthermore
 find permutations as the global data uniquely characterizing the IR fixed point.

\subsection{Network and Moves}\label{sec.bipartitedef}

Let us begin with a (undirected) planar bicolored network $\scG$ (also abbreviated as a plabic network).
Here a bicolored means that the vertices of the graph are colored either black or white.
We will denote the number of 
boundary vertices by $n$, which will be fixed throughout the
analysis below.

\begin{figure}[htbp]
\small
\centering
\includegraphics[width=5cm]{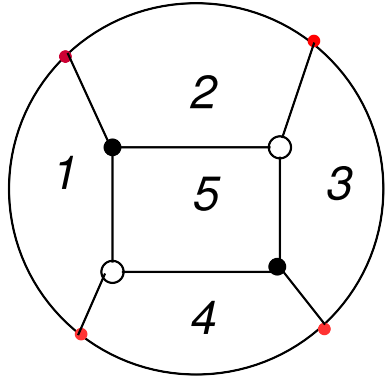}
\caption{An example of a planar bicolored network on a disc with four boundary marked points.}
\label{4net}
\end{figure}

We assume the network is connected and irreducible which means we can not separate the network into the disconnected parts,
 and there are no internal vertices with only one incident edge, we also exclude the case where a loop is attached on a vertex.
This means the network separates the 
disk into faces. The closed faces are inside the disk while the open faces are on the boundary. 
See an example in \Fig \ref{4net}.

\begin{figure}[htbp]
\small
\centering
\includegraphics[scale=0.3]{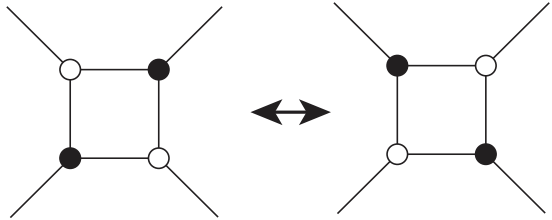}
\caption{Move 1: a square move flips the color of the vertices around a square.}
\label{square1}
\end{figure}

\begin{figure}[htbp]
\centering\includegraphics[scale=0.3]{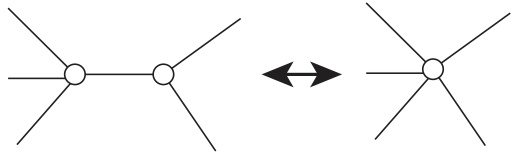}
\caption{Move 2: merging/unmerging of the same-colored vertices does not change the definition of the gauge theory. Similar operation exists for 
black-colored vertices.}
\label{fig.merge}
\end{figure}

\begin{figure}[htbp]
\small
\centering
\includegraphics[scale=.3]{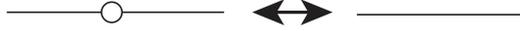}
\caption{Move 3: a degree two vertices can be removed.}
\label{fig.remove}
\end{figure}

\begin{figure}[htbp]
\small
\centering
\includegraphics[scale=0.3]{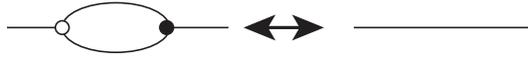}
\caption{Reduction: bubble reduction.}
\label{bubble}
\end{figure}

There are several moves and reductions one can perform on networks.\footnote{Beside the three moves and one reduction discussed in this paper, \cite{PostnikovTotal} introduces two more reductions, leaf reduction and dipole reduction.
These will not, however, play a rule here since we assume that the bicolored graph is connected.} 
The first is the square move (\Fig \ref{square1}). 
The second is to merge or unmerge 
vertices of the same color (\Fig \ref{fig.merge}). 
The third is to remove a degree 2 vertex (\Fig \ref{fig.remove}). 
These three moves do not change the faces of the network. 
Finally, there is the so-called bubble reduction, 
which reduces the number of faces by one (\Fig \ref{bubble}).

A network is called bipartite if all the edges are connecting the vertices
 with different colors.
It is easy to see that one can 
transform any network into a bipartite graph either by merging same-colored vertices or adding degree two vertices. 

\subsection{The Definition of the Theories}\label{sec.networkdef}

We will associate a 4d $\mathcal{N}=1$ quiver gauge theory with any 
connected planar bipartite network $\scG$. 
For a non-bipartite 
network, one could first get a bipartite network by merging the vertices of the same colors.
The rules for the bipartite network are then summarized as follows.

\begin{enumerate}
\item {\bf (gauge and flavor groups):} The network $\scG$ divides the plane into several regions bounded by the edges, each of which we call a face. 
We associate a $SU(N)$ gauge group to each closed face, and a flavor $SU(N)$ group for each open face, i.e. a face at the boundary.

\item 
{\bf (bifundamental matters):}  We associate a $\scN=1$ bifundamental chiral multiplet for each edge except those edges between the white vertices 
and the boundary point.\footnote{This will be necessary for the interpretation of a square move as a Seiberg duality.} The chirality is determined by the black vertices, i.e. the chiral fields form a clockwise direction around it, so the chiral fields 
form a counterclockwise loop around the white vertices.

\item {\bf (superpotential terms):} 
We associate a superpotential term for each simple loop of the quiver diagram, so there is a superpotential term for each vertex except those
white vertices connected to the boundary points.\footnote{Here and in the following we will not be concerned with
the numerical coefficient of the superpotential, as long as it is
generic and non-zero. The change of those coefficients will corresponds to marginal deformations of the IR fixed points, probably along the lines of \cite{Imamura:2007dc}.}

\end{enumerate}

\begin{figure}[htbp]
\small
\centering
\includegraphics[width=10cm]{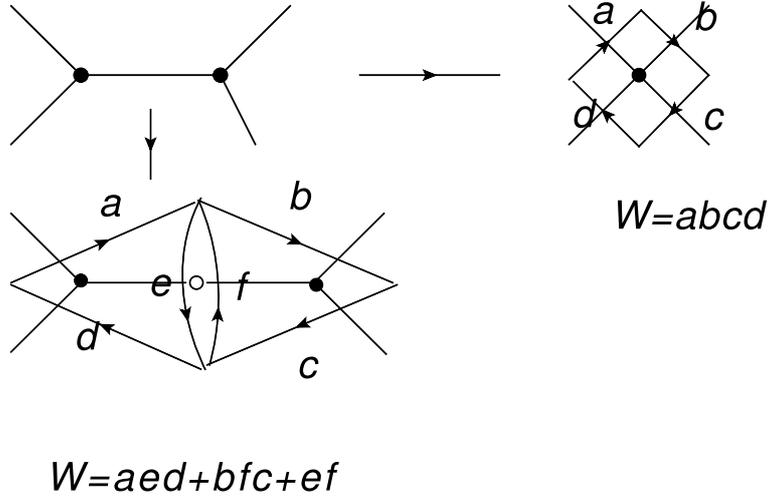}
\caption{The superpotential and matter content for two different ways of making a bipartite graph.}
\label{reduction}
\end{figure}

\begin{figure}[htbp]
\centering\includegraphics[scale=0.35]{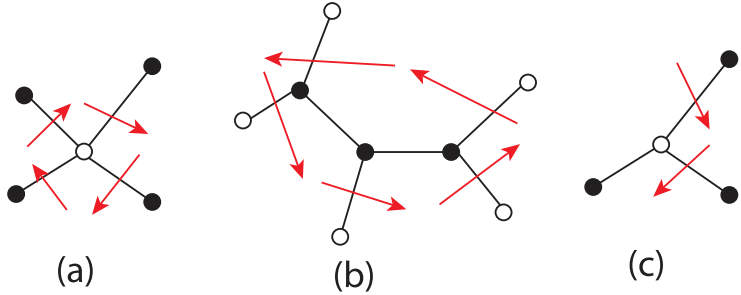}
\caption{Rule for superpotential. We do not include a superpotential term
when the bifundamentals around the vertex does not make a closed loop, as in the case (c).}
\label{fig.Wrule}
\end{figure}

Notice that with the above definition, the theory does not depend on how
many external edges are attached to a white vertex, 
hence we will 
restrict to the case where there is only one external edge for the white vertex without losing anything interesting. 
For a non-bipartite 
graph, instead of merging the vertices of the same colors
we could also get a bipartite graph
by adding degree 2 vertices.
We now show that
these two operations define the same IR fixed point.  The two theories and the corresponding superpotential are given in 
\Fig \ref{reduction}, however, the fields $e$ and $f$ is massive on the left theory and we can integrate out these two 
fields and getting an exactly same matter content and superpotential as the right graph. This shows that
the merging/unmerging operation does not change the IR theory, this also means removing a degree two vertex does not change the IR theory.
With this understanding, the superpotential and the matter fields for any graph is described in \Fig \ref{fig.Wrule}.

Since the number of incoming and outgoing quiver arrows are equal for each face,
and since we assign a rank $N_c=N$ for each gauge or flavor group, 
our quiver gauge theory is anomaly free, and hence is a well-defined theory.
We conjecture that the above defined quiver gauge theory flows to a (potentially trivial) SCFT in the IR.

\bigskip\noindent{\it Seiberg Duality as a Square Move}

We have already seen
that merging/unmerging  of the same-colored vertices  and adding degree two vertices
are trivial operations which keep the IR fixed point.  We would like to translate other two moves into the operations in quantum field theory and show
that they would not change the IR fixed point.
The square move and bubble reduction, which is the most non-trivial move, are identified with a Seiberg dualities, as will be explained below.
One nice match is that for the gauge 
group represented by the square has $N_f=2N$  and $N_f=N$ for the bubble which are the only cases where the 
interesting IR dynamics happen.

\begin{figure}[htbp]
\small
\centering
\includegraphics[width=10cm]{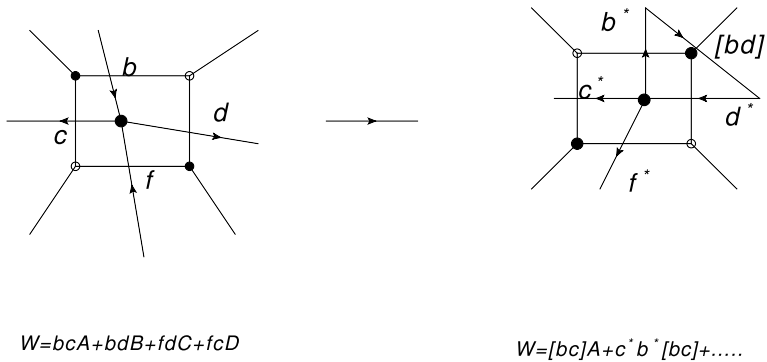}
\caption{A square move of the bipartite graph is identified with a
 Seiberg duality.}
\label{square}
\end{figure}

When $N_f=2 N$, after suitable merging/unmerging, we can assume that the 
vertices around the $SU(N)$ gauge group has alternating colors,
and are of degree 3. Let us start with the local piece of the network and assume the superpotential to have the form depicted in \Fig \ref{square}, where $(A, B, C, D)$ 
represent a sequence of the fields.  If there is only one field in one of the capital letters, that means that the corresponding vertex is degree 3, otherwise
it has degree more than three. The superpotential has the following form
\begin{equation}
W=bcA+bdB+fdC+fcD+....
\end{equation}

After the Seiberg duality,  $(d,b,c,f)$ change chirality, i.e. from fundamental representation to anti-fundamental transformation \cite{Seiberg:1994pq,Berenstein:2002fi}.
Pictorially, the quiver arrows for these fields are reversed, which is in agreement with the square move.
There would be  new mesons $[bd]$,  $[fd]$, $[bc]$ and $[fc]$, and the rank of the new gauge group is still $N$.
The new superpotential is
\begin{equation}
W^{'}=b^*c^*[bc]+b^*d^*[bd]+f^*d^*[fd]+f^*c^*[fc]+[bc]A+[bd]B+[fd]C+[fc]D+....
\end{equation}

This is exactly equivalent to the superpotential obtained from the graph on the right of \Fig \ref{square}. To see this, one might 
need to integrate out the massive fields and simplify the above superpotential if one of the original sequence in $(A, B, C, D)$
has just one field; if one of the sequences, say $A$, has only one field, then $A$ and $[bc]$ are massive, we can integrate them out,  
and set the $A=b^*c^*$ in the superpotential, which is exactly the one given by the  graph after merging the vertices.

When there is a bubble, then the gauge group represented by it has $N_f=N_c$, so there is a mass gap in the IR and the gauge group
is decoupled. The IR physics is nicely interpreted formally as the Seiberg duality where the dual gauge group has rank zero and can be decoupled, and the only low-energy degrees of freedom is the meson field. 
It can be checked the dual theory after decoupling of the gauge group is exactly the same as the one described  by the graph after bubble reduction, and the meson field is represented by an edge of the quiver diagram connecting the two faces.

All the geometric data of the network are mapped perfectly to the gauge theory description, and the moves and reductions are mapped to the gauge theory operations which do not change the IR fixed point.
we summarize 
the mapping in Table \ref{T1}.

\begin{table}[htbp]
\begin{center}
    \begin{tabular}{c|c}
        \hline
       network & gauge theory \\ \hline\hline 
        closed face&gauge group\\  \hline 
        open face&flavor group \\ \hline
         edge& bifundamental matter\\ \hline 
        vertex & superpotential term \\ \hline 
        merging/unmerging & IR theory unchanged\\ \hline 
        adding/removing degree two vertex & IR theory unchanged\\ \hline
        square move & Seiberg duality on $N_f=2N$ \\ \hline 
        bubble reduction & Seiberg duality on $N_f=N$\\ \hline
    \end{tabular}
  \end{center}
    \caption{Dictionary between networks and gauge theories.}
    \label{T1}
\end{table}

\subsection{Reduced Network}
We have defined a $\mathcal{N}=1$ theory for any connected plabic graph. However, the Lagrangian description of the gauge theory could be 
very redundant, i.e. one of the gauge groups could decouple in other duality frames. It would be nice to have some ways to find the minimal duality frame, 
where we could not further reduce the number of gauge groups and matter fields. Since the gauge theory operations which do not change the IR 
fixed point have been mapped to geometric operations on the network, the
above-mentioned task is translated to a combinatorial
study of the network, which has been 
studied in full detail in the mathematical literature.

\begin{figure}[htbp]
\small
\centering
\includegraphics[width=6cm]{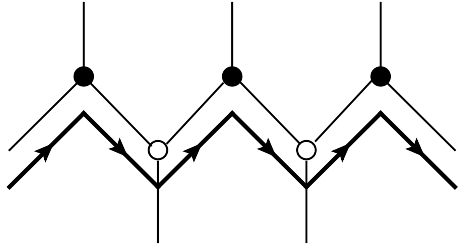}
\caption{A zig-zag path which is turning left at white vertex and turning right at the black vertex.}
\label{zig}
\end{figure}

We call a network reduced\footnote{This is also called minimal in the literature.} when one cannot reduce the number of faces by the four moves/reduction above;
the gauge theory description corresponding to this type of network is exactly the minimally duality frame we want to find.
It seems hard to judge whether a network is reduced or not from this definition,  i.e. 
to judge whether a complicated quiver gauge theory to 
be reduced or not. 
Fortunately there is an easy combinatorial algorithm to achieve this task on the network,
and therefore we have a nice way of finding the minimal duality frame of a quiver gauge theory.
Before describing the method, one 
needs to introduce two objects on our network.  
First, a zig-zag path\footnote{This is called a trip in \cite{PostnikovTotal}, and is also called a train track in the literature.} is a path on the network
 which turns maximally left at white vertex and turns maximally
right at black vertex (\Fig \ref{zig}). 
These paths come in from boundary points, and again go out to boundary points after several turns.

\begin{figure}[htbp]
\small
\centering
\includegraphics[width=8cm]{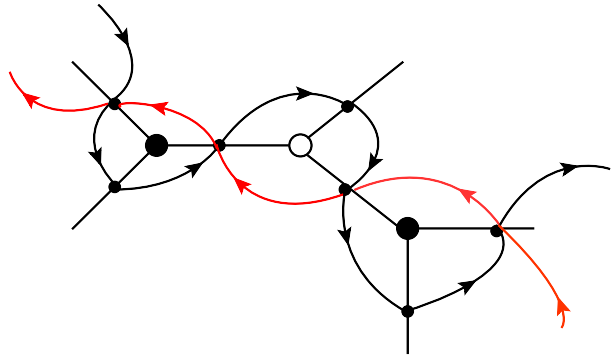}
\caption{A strand for a bipartite network.  Put one vertex on each edge and connect these dots around each vertex. The orientation is taken to clockwise for the white vertex and 
counterclockwise direction for the black vertex.}
\label{strand}
\end{figure}

The second is a strand.\footnote{In some literature a strand is meant to be a zig-zag path in the definition here.}
Place one vertex on each internal edge and connect the dots around each vertex. Take the 
clockwise orientation for segments around the white vertex and counterclockwise orientation for the black vertex, see \Fig \ref{strand} for illustration. The network is 
replaced by another graph with all vertices degree four. A strand is defined as an oriented path such that it represents the zig-zag path (see \Fig \ref{strand}).

\begin{figure}[htbp]
\small
\centering
\includegraphics[width=6cm]{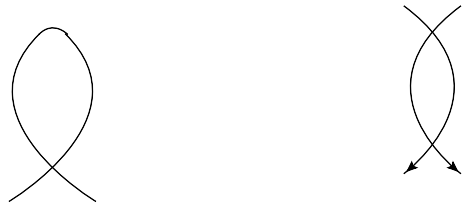}
\caption{The bad configurations for the strand, forbidden in reduced bicolored graphs.}
\label{bad}
\end{figure}

Now we can introduce the criteria for a network to be reduced \cite[section 13]{PostnikovTotal}: there is no bad configuration of strands as described in \Fig \ref{bad}; moreover, there is 
no closed zig-zag loop in our network.  If there are bad strand configurations, then one can find a bubble after a sequence of the moves, notice that the strand structure is only changed 
by bubble reduction. There is a canonical way of finding and  eliminating bubbles. We draw the strand on the glued network and do square moves on each  bounded area  of 
the bad strand configuration shown in the \Fig \ref{bad}, bubbles will appear at some stage and we eliminate it using bubble reduction. 
The network become minimal after eliminating all the bubbles in this way. 
A useful property of a reduced planar bicolored network, when the network is connected, is that we have
\beq
V-E+F=1 \ ,
\eeq
where $V, E, F$ are the number of internal vertices, edges and faces of $\scG$, respectively.
In the example of \Fig \ref{4net} we have $V=4, E=8, F=5$.

\begin{figure}[htbp]
\small
\centering
\includegraphics[width=10cm]{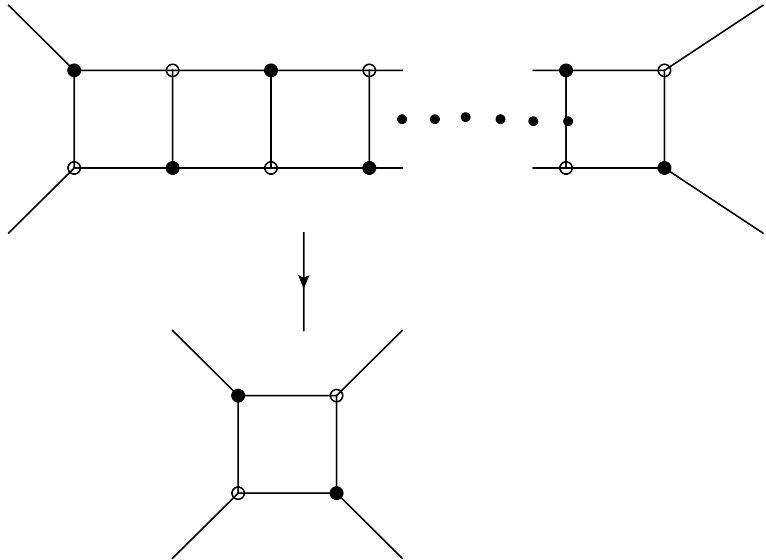}
\caption{By combining the moves and the bubble reduction, we could sometimes simplify a bicolored graph considerably.}
\label{hugereduction}
\end{figure}

The reduction is rather powerful, and a complicated network may be substantially simplified, see an example in \Fig \ref{hugereduction}. This result is 
rather remarkable from the gauge theory point of view, since 
the initial
gauge theory with a huge number of gauge groups
reduces to a single gauge group in the minimal duality frame!

\subsection{Classification by Permutations}\label{sec.permutations}

We have described how to find a minimal duality frame for any quiver gauge theory defined by the network: it corresponds 
to the reduced network.
However, the minimal duality frame is not unique, it might be possible that there are many minimal duality frames.
It would be really nice if there is a way to judge whether two minimal quiver gauge theories describe the same IR fixed point. 

What we want is a global characterization of the 
IR fixed points, as emphasized in introduction;
we need an algorithm to determine, given two reduced planar bicolored networks $\scG$ and $\scG'$,
whether the two could be related by a sequence square moves. 
In other words we would like to extract some global 
data
from a reduced planar bicolored graph such that two graphs are related by the moves if and only if these data match.
Fortunately, there is already an answer in mathematics literature \cite{PostnikovTotal}, which we will now explain.

Let us first label the boundary points from $1$ to $n$ (this is defined modulo translation ambiguities).
Then each zig-zag path (strand) starting from point $i$ will reach another point $j$.
Note that this is a one-to-one map. For example, if two paths from $i$
and $j$ end up the same point $k$, we can go backwards in the path $k$
and find that $i=j$.
This means that 
our network $\scG$ defines a permutation, which we denote by $\pi_{\scG}$.\footnote{Actually in \cite{PostnikovTotal} this theorem is stated for 
a decorated permutation, not an ordinary permutation.
Here a decorated permutation is a pair $\pi \in \mathfrak{S}_n$
together with a coloring function $\textrm{col}$ 
from the fixed point set $\{ i| \pi(i)=i\}$
to $\{ 1, -1\}$. 
The coloring is chosen such that 
if the strand for the 
zigzag path is a counterclockwise loop (reps., clockwise) loop, then the fixed point $i$ is colored in black ${\rm col}(i)=1$ (reps., white and ${\rm col}(i)=-1$). Such a fixed point is associated with a lollipop in the bicolored network, but we have assumed already that this does not happen (recall section \ref{sec.bipartitedef}), and hence we can disregard the decoration in this paper.
\label{footnote.decorated}
}

\begin{figure}[htbp]
\centering{\includegraphics[scale=0.45]{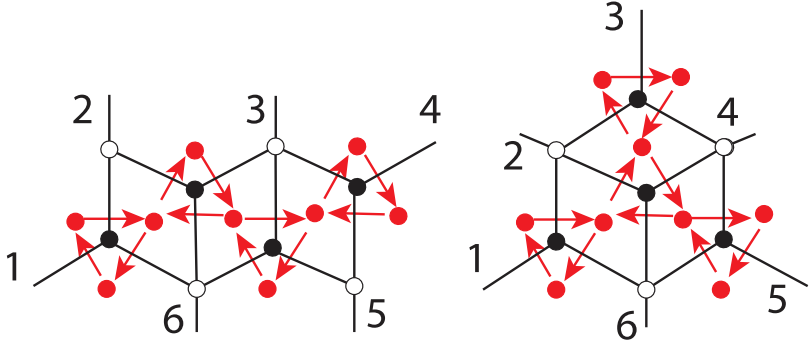}}
\caption{The two bipartite graphs are related by a sequence of moves,
and define the same IR fixed point. Correspondingly, 
the two graphs give the same permutation $\pi: \{ 1, 2, 3, 4, 5, 6\}\to \{3, 4, 5, 6, 1,
 2\}$.}
\label{fig.permeg}
\end{figure}

Now we have the following theorem (\cite[Theorem 13.4]{PostnikovTotal}),
which provides a very simple criterion to judge whether two quiver gauge theory associated with two reduced network flow to 
the same IR fixed point:
suppose we have two reduced planar bicolored networks $\scG, \scG'$ with the same number of 
boundary vertices. Then $\scG'$ is obtained by a sequence of square moves from $\scG$ 
if and only if these two networks have the same permutation (up to a cyclic shift):
$\pi_{\scG}=\pi_{\scG'}$ (see \Fig \ref{fig.permeg} for an example).\footnote{
It is important that $\scG, \scG'$ are reduced here, since bubble
reduction \ref{bubble} changes the permutation.}
This classification is quite powerful, since there are only a finite possible number of permutation of $n$
if we fix $n$.\footnote{There are many more possibilities if we lift the condition that the planar network is connected (see also footnote \ref{footnote.decorated}).}

\begin{figure}[htbp]
\small
\centering
\includegraphics[scale=0.3]{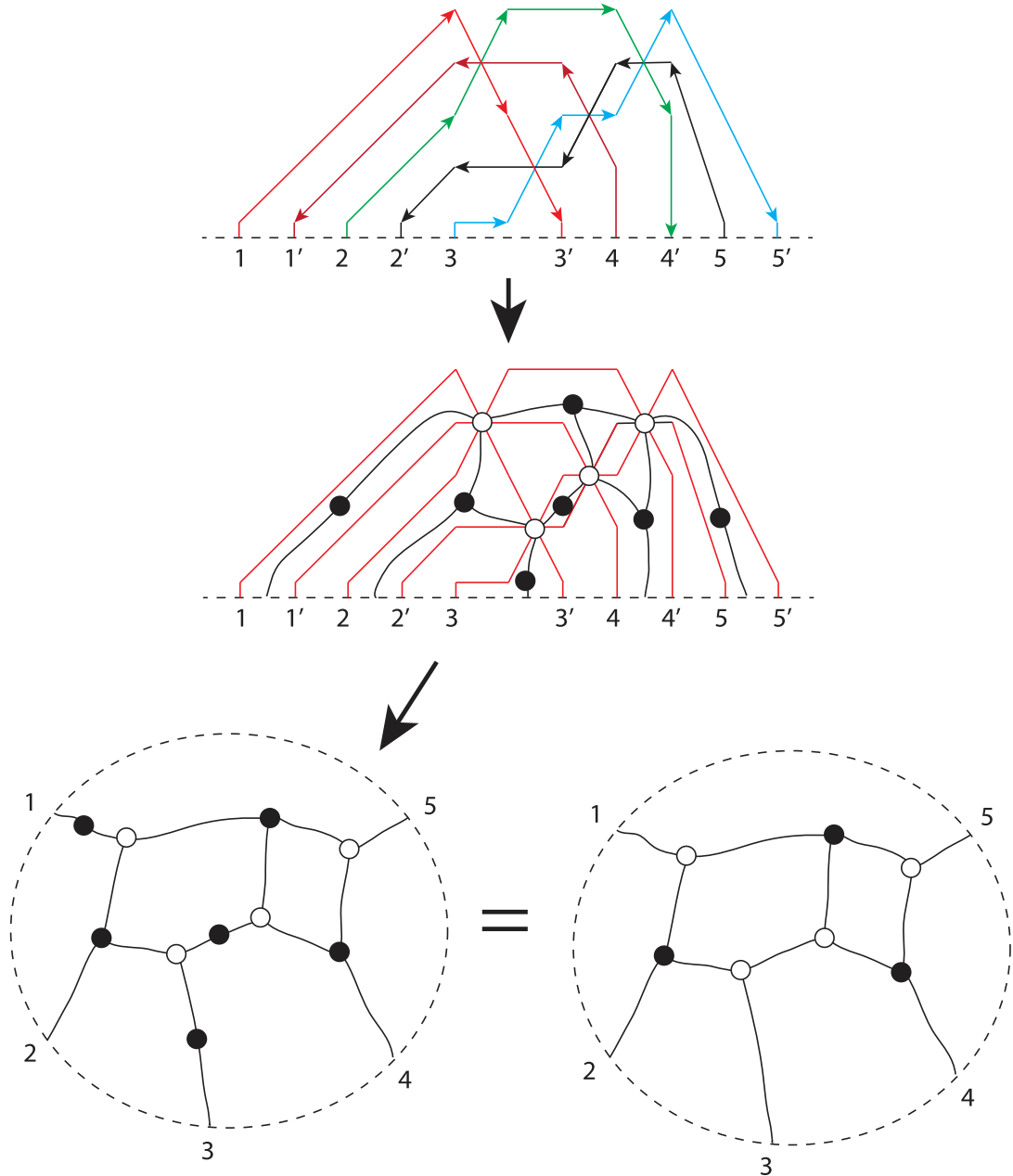}
\caption{The inverse map, constructing a planar bicolored network from a permutation.}
\label{fig.inverse}
\end{figure}

There is an inverse to this construction; not only can one identify the permutation from a given network, but
we can also construct a network from a  given permutation (see \cite{ThurstonDomino} and \cite[section 6]{PostnikovTotal}).
We first write $1, 1', 2, 2',\ldots, n, n'$ along the boundary of the disc.
We then draw paths from  $i$ to $\pi(i)'$
such that the paths always intersect at a triple intersection point.
We can then place white vertices at the triple intersection points,
and black vertices to each region surrounded by the paths in a clockwise orientation (\Fig \ref{fig.inverse}).

\subsection{Totally Non-negative Grassmannian}\label{sec.Gr}

We have seen that our theory is defined from a (decorated) permutation.
Interestingly, it is known that such an object 
is in one-to-one correspondence 
with a cell of the totally non-negative Grassmannian.

Recall that a Grassmannian manifold $\Gr_{k,n}:=\Gr_{k,n}(\bR)$ is the space of 
$k$-dimensional vector spaces inside $\bR^n$. We define the non-negative Grassmannian 
$(\Gr_{k,n})_{\ge 0}$ as the subspace where the Pl\"{u}cker coordinates are all non-negative.
This space $(\Gr_{k,n})_{\ge 0}$ has a cell decomposition, 
given by non-negative parts of the matroid strata.

The decorated permutation discussed in this paper is in one-to-one correspondence with a 
cell of this non-negative Grassmannian $(\Gr_{k,n})_{\ge 0}$.
Here that $n$ is the number of boundary punctures as defined previously, 
and an integer $k$ is defined by
\beq
k-(n-k)=\sum {\rm col}(v)({\rm deg}(v)-2) \ ,
\label{kdef}
\eeq
where ${\rm col}(v)=1$ for black vertices and ${\rm col}(v)=-1$ for white vertices.
We can verify that this integer $k$ does not change
under the moves, and hence is a well-defined
data assigned to the IR fixed point.
The planar examples in the previous section is the case $k=2$.
The Grassmannian duality $k\to N-k$ is translated into the flip of the 
color of the vertices.

The moves introduced earlier does not change the cell it describes, similarly, the moves 
do not change the IR fixed point of the four dimensional gauge theory. So there is 
a correspondence between the positive cells of the Grassmannian to a four dimensional $\mathcal{N}=1$
gauge theory. This has some remarkable applications, for example, whenever we have 
a network with $k=1$, the reduced gauge theory is trivial! This is because the only
reduced connected network describing a cell of $(\Gr_{1,n})_{\ge 0}$ has only one white vertex with all the boundary points connected to it.

Interestingly, exactly the same mathematical structure of the totally non-negative Grassmannian, 
appears in the recent work on scattering amplitudes \cite{ABCGPT}. 
There are many formal similarities; for example, the black/white color in our story represents the chirality 
there. It remains to be seen, however,
if there is a deeper reason why the same mathematical structure appears
in two different physical setups.

\subsection{Comments on Non-Planar Networks}\label{sec.highergenus}

The definition in this section has so far been limited to the planar networks.
This raises an obvious question: can we generalize our definition
to bicolored networks on more general bordered Riemann surfaces?
Part of this will be discussed in section \ref{sec.def1}, 
but the question here is about a generalization to bicolored graphs not 
necessarily coming from a triangulation.

\begin{figure}[htbp]
\small
\centering
\includegraphics[scale=0.3]{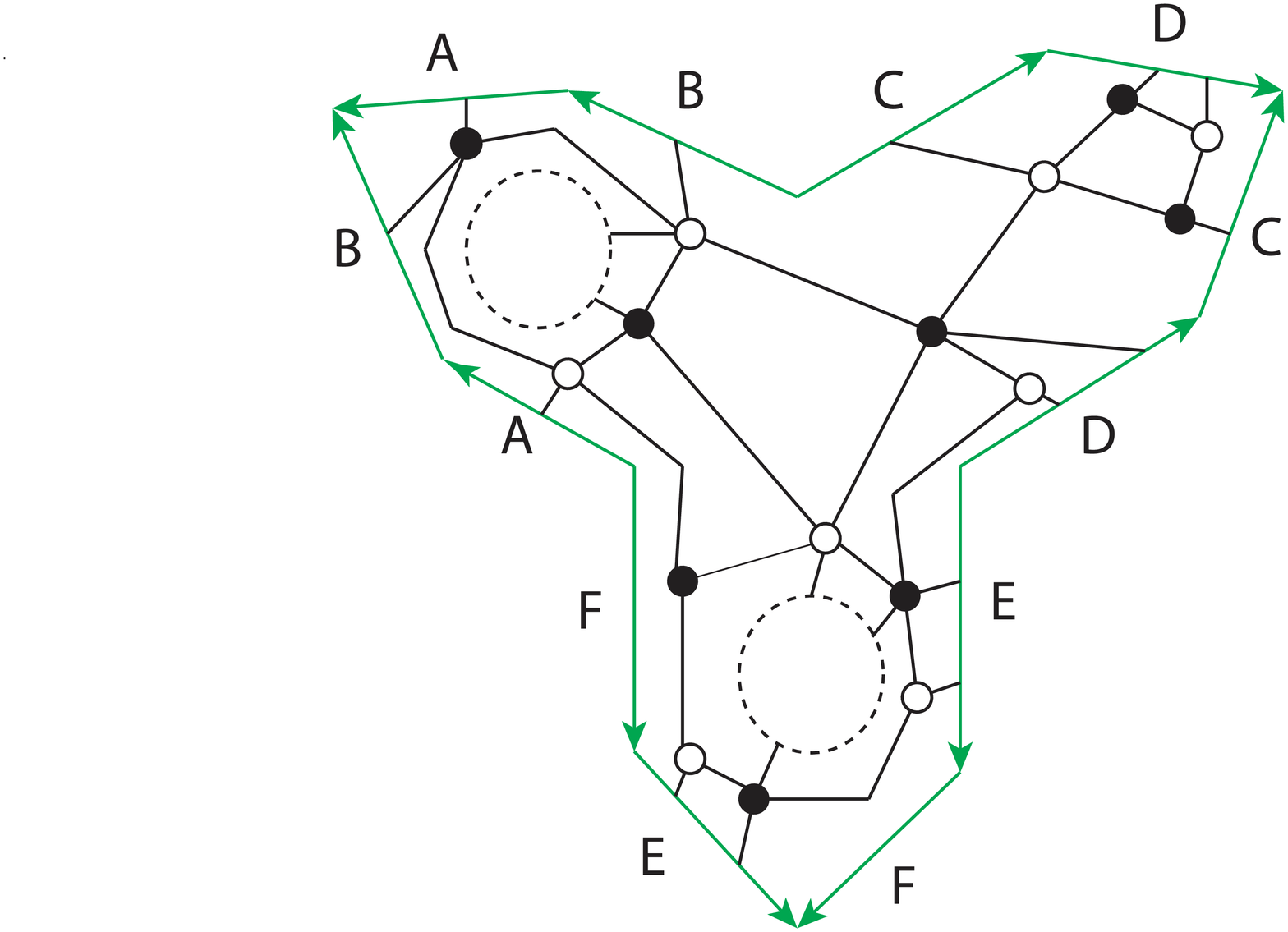}
\caption{An example of network on a genus $3$ surface with $2$ boundaries components.}
\label{fig.genus2}
\end{figure}

It is actually straightforward to adopt the planar definition to the 
more general case.
Let us start with a bipartite network on a bordered Riemann surface.
Some of the edges could run off to infinity at the boundary,
however we impose the condition that they pass through the 
marked points at the boundary (see \Fig \ref{fig.genus2}).
We also impose the conditions that the bipartite graph is connected,
and that there are no degree $1$ internal edges. 
The graph then divides the Riemann surface into contractible faces.

We find again that a network divides the surface into faces,
each of which is topologically a disc, i.e. all the faces are contractible.
We can then use exactly the same rules in 
section \ref{sec.networkdef} to define our 4d $\scN=1$ theories.
Note that the rules are local on the network, and hence does not require the global information 
of the surface.

While this definition is straightforward,
what is missing here is the global data 
uniquely characterizing the IR fixed point,
the counterpart of a decorated permutation. 
We will also need to generalize
the definition of reduced graphs to non-planar networks, and 
prove a non-planar generalization of \cite[Theorem 13.4]{PostnikovTotal}.

We do not give an answer to this problem in general,
and would like to leave it for future work.
Instead we here point out that the answer to this 
question is known, when the Riemann surface is a torus 
without any boundaries.
The theories in this case have extensively been 
studied in the context of 4d $\scN=1$ quiver gauge theories
dual to toric Calabi-Yau manifolds, 
and the bipartite graph in this context is called a brane tiling (see \cite{Hanany:2005ve,Franco:2005rj,Franco:2005sm}
and the reviews \cite{Kennaway:2007tq,Yamazaki:2008bt}).\footnote{The name ``brane tiling'' is justified by the
fact that the bipartite graph is a certain reduction of an actual brane configuration, see \cite{Yamazaki:2008bt} for detailed analysis.}

The definition of a zig-zag path is parallel 
to the planar case. Since the surface does not have any boundary components,
instead of starting with a marked point on the boundary we start with an arbitrary vertex on the 
graph and define the zig-zag paths by following the same rule as
in the planar case (turns maximally right/left at black/white vertex).
Sine the graph is finite, the path always will be a closed loop.
By counting the winding numbers in some basis of $H_1(\Sigma)$ (say, the $\alpha$-cycle and the $\beta$-cycle),
we have a set of integers $(n_{1,i}, n_{2,i})$ for each zig-zag path $p_i$.
The convex polytope for these lattice points in $\bZ^2$ is identified with the toric diagram,
which in turn defines the geometry of the toric Calabi-Yau manifold 
dual to these theories \cite{Hanany:2005ve,Franco:2005rj}.
Here the choice of the basis of $H_1(\Sigma)$ is translated into the $SL(2, \bZ)$ ambiguity of the torus.
This convex polytope could be thought of as a torus counterpart of a decorated permutation.
There is also a inverse map, to obtain a bipartite graph from a toric diagram.
This has been worked out in \cite{Hanany:2005ss}, and (in the mathematical reformulation of \cite{Goncharov:2011hp})
uses essentially the same ingredient as the procedure in \Fig \ref{fig.inverse}.

Moreover, under a suitable definition of a reduced graph
we can show that two reduced planar bipartite graphs are connected by a sequence of the three moves in section \ref{sec.bipartitedef}
if and only if they correspond to the same convex polygon (up to $SL(2, \bZ)$-transformation and translation in $\bZ^2$), see \cite[Theorem 2.5]{Goncharov:2011hp}.

We can therefore regard the theories specified by bipartite graphs on $T^2$
as a particular example of more general theories associated with a
bicolored network on a bordered Riemann surface.\footnote{See \cite{Hanany:2012vc}
for a recent discussion on 4d $\scN=1$ gauge theories associated with a bipartite graph on a higher genus Riemann surface.
}

For a general bordered Riemann surface, 
we conjecture that the global data are a combination of 
(decorated) permutations associated with the boundary 
marked points, as well as the set of integers
counting the winding around the non-trivial cycles
of the Riemann surface. 
It remains to be an exciting problem to find out if this is really the case,
and if these data fit nicely into a 
 geometrical object generalizing the totally non-negative 
Grassmannian of \cite{PostnikovTotal}.

In next section, we are going to study a subclass of network defined on higher genus surface with boundaries.
Instead of starting with a random network, we are going to begin with some global data  from the definition of
the Riemann surface. This class of theories have the exactly same properties as those arising from
the planar network.

\section{Bordered Riemann Surface and Network}\label{sec.def1}
The goal of this section is to extend the definition of a network to higher genus bordered Riemann surface with marked point.
We are going to  define a 4d $\scN=1$ theory $\scT_{\Sigma}^l$
for each bordered Riemann surface,
for each fixed integer $l$. Although not quite general as the planar network, this class of network is reduced and built from
the basic components, therefore they give a nice way of constructing reduced network even in the planar case.

\begin{figure}[htbp]
\centering\includegraphics[scale=0.3]{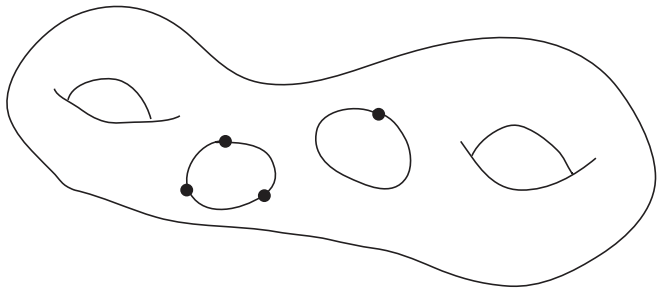}
\caption{Our $\Sigma$ is a bordered Riemann surface, whose boundary
 consists of disjoint union of circles. On each circle we specify at
 least one marked point.}
\label{fig.marked}
\end{figure}

Let us begin with a Riemann surface with boundaries,
and specify a finite set of points $M_{\rm boundary}$,
called boundary marked points,
on the boundary circles of $\Sigma$.
Each connected component of $\partial \Sigma$ has at least one
boundary marked point; we will explain later why we exclude the boundary without marked point.
The defining data of our theory is a pair
$(\Sigma, M_{\rm boundary})$, 
see \Fig \ref{fig.marked} for an example.
For notational convenience we
sometimes denote this pair simply by $\Sigma$. 
In other words, $\Sigma$ is defined by following data:

a. the genus $g$ of the Riemann surface;

b. the number $b$ of boundary components;

c. the number of marked points $h_i$ on each boundary.

\noindent We require there are at least three marked points for the disc, and at least one boundary component for higher genus Riemann surfaces.

Instead of directly constructing a network, we are going to define two nice equivalent combinatoric 
objects on a bordered Riemann surface: open pants decomposition and ideal triangulation. 
We are going to define $\scN=1$ theories from them.
The combinatoric objects are not unique but are related by a sequence of local moves: $s$-$t$ duality 
of pants decomposition and a flip of the triangulation, both of which are equivalent to the square move introduced in previous section.
This ensures that different Lagrangian 
descriptions are related by a sequence of Seiberg dualities, and therefore define the same IR fixed point. In the end, 
an equivalent bipartite network is constructed from the ideal triangulation which justifies that they are just 
a subclass of the theories from the network.

\subsection{Open Pants Decomposition}\label{sec.open}

This definition is in analogy of the Gaiotto's interpretation of S-duality of $\mathcal{N}=2$ theory \cite{Gaiotto:2009we}.  
The SCFT is defined by a bordered Riemann surface with marked points
 and there is a global $SU(N)$ flavor symmetry associated with each marked point.  We are interested in the UV Lagrangian description of the IR fixed point.
In this section we focus on the case $g=0$, and begin with a disc with several marked points. The Lagrangian description is derived by looking at 
the degeneration limit of the Riemann surface into three punctured discs.  New marked points appear in the complete degeneration limit.
Physically, each three punctured disc represents a 
matter system with $SU(N)^3$ flavor symmetry and the gluing of the Riemann surface is interpreted as 
gauging the diagonal flavor symmetry  associated with the marked points.  Our proposal is that the $\mathcal{N}=1$ Seiberg duality corresponds
to different degenerations of the disc.  

The first step for the definition of the Lagrangian is to associate a
theory $\scT_{\Delta}^l$ to 
each three punctured disc $\Delta$.
We first concentrate on the case $l=1$.
We propose that this theory is a 4d $\scN=1$ theory with 
global symmetry $SU(N)^3$, together with
three $\scN=1$ bifundamental chiral multiplets
$X, Y, Z$ transforming in the representation 
$(N, \overline{N}, 1)$, $(1, N, \overline{N})$, $(\overline{N}, 1,N)$
under the  $SU(N)^3$ global symmetry.
We also include a superpotential term $W=\textrm{Tr} (X Y
Z)$.

\begin{figure}[htbp]
\centering\includegraphics[scale=0.4]{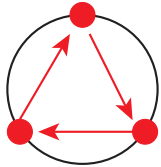}
\caption{A quiver diagram for the triangle theory $\scT_{\Delta}^{l=1}$.}
\label{fig.trianglequiver}
\end{figure}

It is useful to represent this by a quiver diagram in \Fig \ref{fig.trianglequiver}.
Note that the theory is kept invariant under the cyclic rotation of the triangle;
we simply need to change the label of bifundamental chirals.
However, since the theory is chiral (except for the special case $N=2$)
we do not the the symmetry under the orientation reversal of the
triangle. 

\begin{figure}[htbp]
\centering{\includegraphics[scale=0.5]{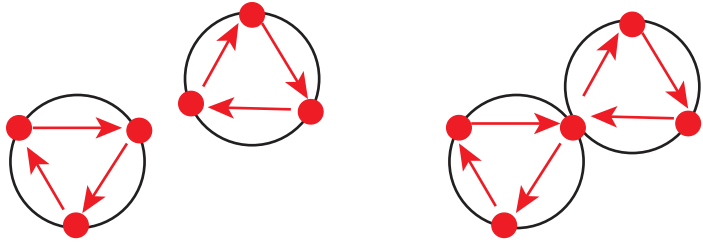}}
\caption{The theory for the two triangles glued along an edge.}
\label{fig.triangleglue}
\end{figure}

The second step is to glue two three punctured discs (or equivalently triangles). 
Let us glue two triangles $\Delta_1, \Delta_2$ 
along the edge $e$, and  
define the corresponding theory $\scT_{\Delta_1 \cup_e \Delta_2}$
(see \Fig \ref{fig.triangleglue}).
The edge $e$ corresponds to global symmetries $SU(N)_1, SU(N)_2$
for the theories $\scT_{\Delta_1}, \scT_{\Delta_2}$,
and by gauging the diagonal $SU(N)$ in $SU(N)_1\times SU(N)_2$
we obtain a $SU(N)$ gauge theory with six bifundamental 
chiral multiplets $X_i, Y_i, Z_i$ ($i=1,2$).
This theory has a superpotential 
$W=\Tr(X_1 Y_1 Z_1+X_2 Y_2 Z_2)$, and has 
$SU(N)^4$ global symmetries.
At the level of the quiver diagram, this gluing is simply to 
identify the two vertices of the quiver diagram corresponding to the 
marked point $e$. 

The gluing of more three-punctured discs is similar, and there is one Lagrangian description for 
each degeneration. By construction the number of 
fundamentals and the anti-fundamentals are the same for all $SU(N)$
gauge groups, and hence 
the chiral anomaly automatically vanishes for our theory $\scT_{\Sigma}$.

We now conjecture (at least for a generic cases)
these theories flow to a non-trivial IR fixed point,
and 
we will primarily be interested in this strongly coupled theory in IR.

\begin{figure}[htbp]
\small
\centering
\includegraphics[width=10cm]{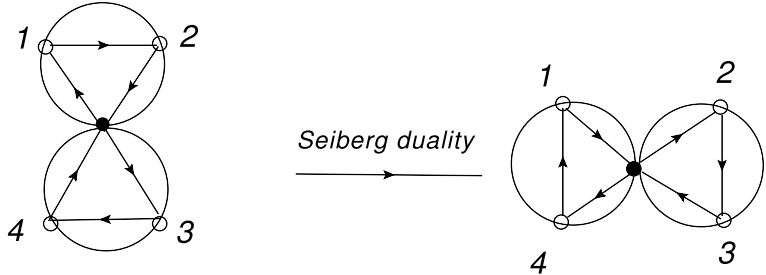}
\caption{Two different degeneration limits correspond to two duality frame of the same theory.}
\label{disk}
\end{figure}

The basic Seiberg duality corresponds to the fourth punctured disc as shown in
\Fig \ref{disk}. Since there are only two degeneration limits ($s, t$-channels) to preserve the cyclic
order, there are only two duality frames. The above interpretation is different from the $\mathcal{N}=2$
duality proposed in \cite{Gaiotto:2009we}, where there is no cyclic order for the four
punctures and we have three duality frames ($s, t, u$-channels).

\begin{figure}[htbp]
\small
\centering
\includegraphics[width=10cm]{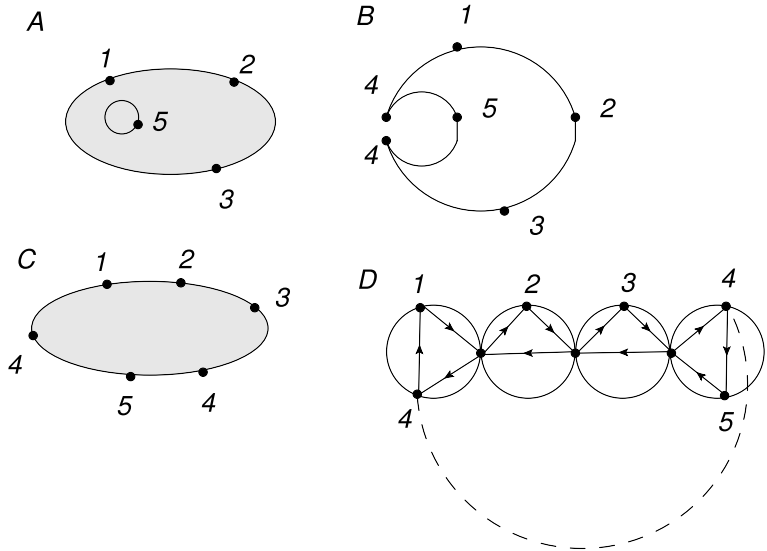}
\caption{A: a Riemann surface with two boundaries and marked points, one boundary is interpreted as a hole in the disc.  B: The degeneration of the hole. C:
the disc with a cyclic order of marked points after the degeneration. D: One duality frame of the theory from A.}
\label{holedegeneration}
\end{figure}

We can also add another boundary with marked points to the Riemann surface, and there are two steps in studying the degenerations 
(see \Fig \ref{holedegeneration}). First, there are two new marked points with the same label appearing in 
degenerating the hole and the Riemann surface becomes a disc, and one can find the theories in any duality frame according to the rule described earlier. We then gauge 
the diagonal flavor symmetries associated with the newly-appearing punctures.  
Following similar procedure, one can find the theory for the disc with arbitrary number of holes.

One could also consider theories from higher genus Riemann surface using the degeneration and gluing idea, however,
this approach is not that illuminating and we are going to propose another easy method in next subsection.

\subsubsection{Comparison with Gaiotto Theories}

\begin{figure}[htbp]
\small
\centering
\includegraphics[width=12cm]{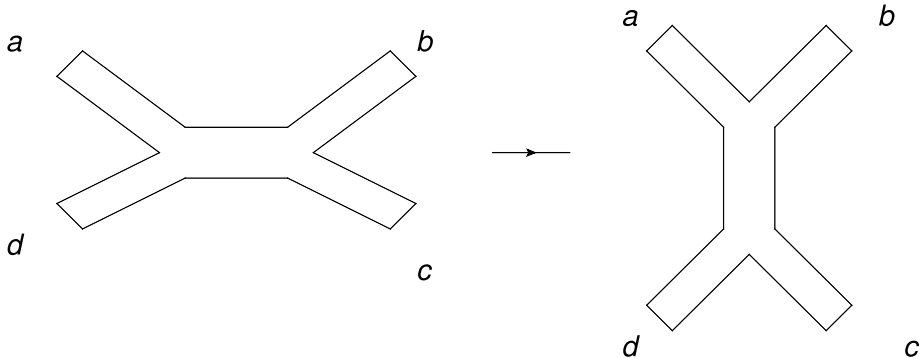}
\caption{Open pants decomposition a disc with four marked points has two duality frames.}
\label{openpants}
\end{figure}

As already mentioned, our definition of $\scT_{\Sigma}$ follows the 
same philosophy as in the 4d $\scN=2$ theories defined by
compactification of M5-branes on Riemann surfaces \cite{Gaiotto:2009we}.
There we associate a theory $\scT_N$ to a trinion (three-punctured
sphere),
and given a pants decomposition (degeneration) of the Riemann surface 
the theory is defined 
by gluing $\scT_N$ theories by gauging diagonal global
symmetries. Different pants decompositions correspond to 
different duality frames of the gauge theory. 
Our theory follows exactly the same step, where we used open pants instead of 
closed pants (see \Fig \ref{openpants}).

Despite the similarities it is important to keep in mind several crucial
differences between (1) our theories and (2) Gaiotto theories.
First, the two theories do not agree even if we take the same Riemann
surface;
Gaiotto theories have 4d $\scN=2$, whereas our theories have 4d
$\scN=1$, and our theory for the 3-punctured sphere $\scT_{\Delta}$ is different from
the $\scT_N$ theory. Related to this is that gluing involves gauging of 
a 4d $\scN=1$ vectormultiplet in (1) and $\scN=2$ vectormultiplet in
(2). The two also uses different data in the definition; in (1) we have
an integer $l$ (as explained in next subsection) and another integer $N$ (specifying the rank of the gauge
group), whereas in (1) we have a single integer $N$ specifying the number of
M5-branes.
Finally, Gaiotto theories are defined from compactification of
M5-branes,
whereas in our case there is apparently no such direct brane
interpretation; our theory is defined purely from the 
combinatorial data of the triangulation.

To some readers the fact that our theories does not have (at least as of this writing)
direct brane interpretation might seem like a disadvantage of our
approach.
However, as already emphasized in Introduction, the idea of 
understanding the global structure of duality symmetries in
supersymmetric gauge theories from geometric data is more general 
than the compactification of M5-brane theories, and our theories could
be thought of as one of the first examples of this sort.

\subsubsection{Generalizations}
\begin{figure}[htbp]
\centering\includegraphics[scale=0.45]{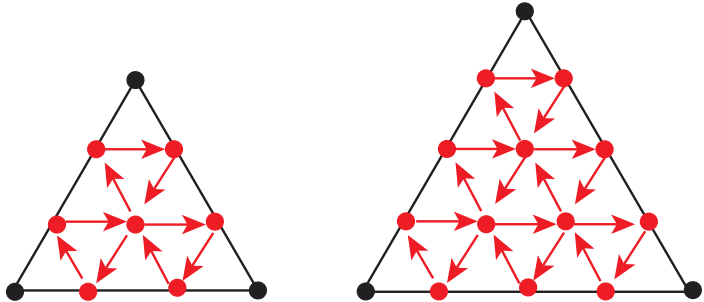}
\caption{The theory $\scT_{\Delta}^{l}$ for $l=2,3$. There is a superpotential term for each cyclic triangle.}
\label{fig.An}
\end{figure}

A natural generalization is to replace the three punctured theory $\scT_{\Delta}$ such that the two theories associated with the different degenerations are related by Seiberg duality.
Motivated by the seminal paper on higher Teichmuller theory by Fock and Goncharov \cite{FockGoncharovHigher},\footnote{
This theory has nice application to 3d hyperbolic geometry and 
3d $\scN=2$ gauge theories \cite{DGXY}.
} 
we introduce the $\scT_{\Delta}^l$ theory show in \Fig \ref{fig.An}.
We introduce superpotential terms for 
any cubic loop of the quiver diagram. For example, the triangle theory $\scT_{\Delta}$
itself has $l(l-1)/2$ superpotential terms.

\begin{figure}[htbp]
\centering\includegraphics[scale=0.45]{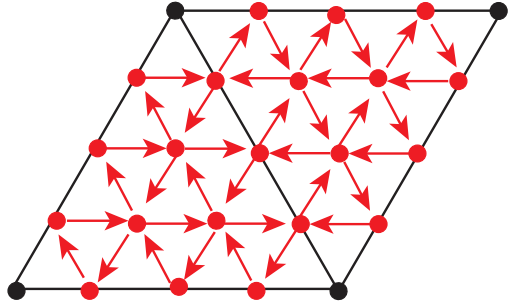}
\caption{The theory $\scT_{\Delta}^{M=3}$ for a disc with four
 punctured is obtained by gluing two triangle theories $\scT_{\Delta}^{M=3}$.}
\label{fig.Anglue}
\end{figure}

There are $l$ $SU(N)$ flavor symmetries per each marked point, 
and the gluing of the marked points is achieved by gauging these $l$ symmetries.  We also 
need to include further superpotential terms for the newly formed simple loop resulting from the gauging.
It is a non-trivial fact
that the theory associated with two different degenerations of the fourth punctured disc are related by Seiberg duality. The special assignment of the superpotential 
is crucial for this fact.
Let us check this for $l=2$ (\Fig \ref{fig.Anglue}).  
In gauging 
two three punctured discs, 
there would be a new superpotential term for the newly formed cycle.  Now let's first do Seiberg duality on node $a$ and $b$, since they have 
$N_f=2N$. After integrating out the massive fields, the gauge group $c$ and $d$ now has flavor $N_f=2N$ and we can do Seiberg duality on these nodes, and the final 
dual theory is exactly the same as the one defined by another degeneration limit of the fourth punctured disc. The general case can be found in \cite{Xie:2012dw,GoncharovIdeal}. 
\begin{figure}[htbp]
\small
\centering
\includegraphics[width=10cm]{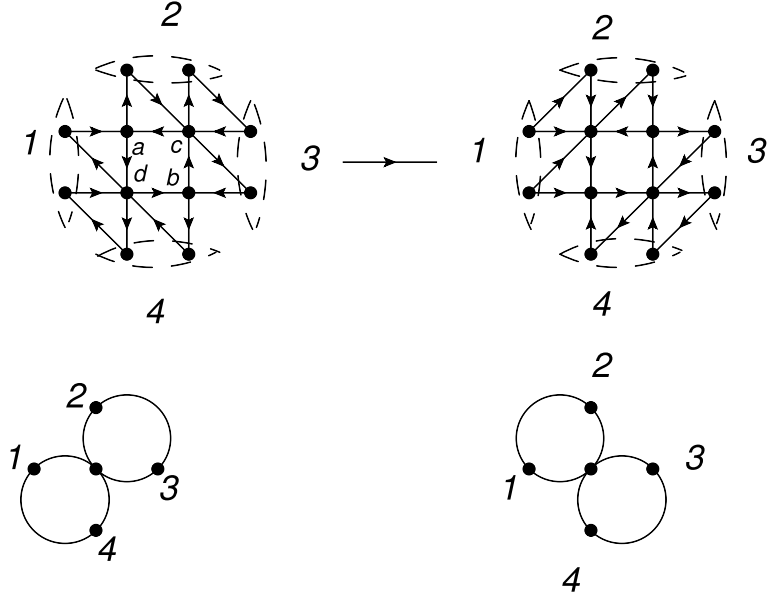}
\caption{One first do Seiberg duality on node $a$ and $b$, and then do Seiberg duality on node $c$ and $d$, the dual quiver is the same as the one defined by another 
degeneration limit.}
\label{gluing}
\end{figure}

In the description above all the marked points are identical, however we can generalize this and use different punctures labeled by rank $l$ Young tableaux. The corresponding three punctured 
disc theory and the Seiberg duality are described in \cite{Xie:2012dw,GoncharovIdeal}.

\subsection{Ideal Triangulation}\label{sec.triangulations}
Now let us extend the definition of the theory to higher genus surface.
This definition will be equivalent to the previous one if 
the genus is zero. Definition of our theory involves a triangulation of the surface,
however we show that theories defined from different triangulations are 
related by a sequence of local operations called a flip. 
The flip matches perfectly with the basic Seiberg duality,
hence they define a unique IR fixed point 

To define our theory we need to fix an ideal triangulation of $\Sigma$.  An ideal triangulation 
is defined using arcs \cite{FST1}.  A simple arc $\gamma$ in $\Sigma$ is a curve such that 

a. the endpoints of $\gamma$ are marked points; 

b. $\gamma$ does not intersect itself, except at the endpoints;

c. $\gamma$ is disjoint from the marked points and the boundary.

\begin{figure}[htbp]
\centering\includegraphics[scale=0.3]{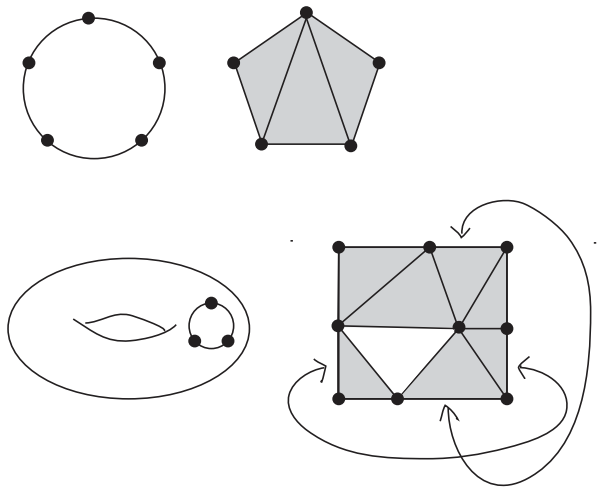}
\caption{A bordered Riemann surface with marked points (a disc with five marked points and
 a torus with three marked points), its triangulation
and the dual graph. The triangles are represented by gray regions, and 
there is no triangle in the white region.}
\label{fig.disc}
\end{figure}

We also require the arc $\gamma$ is not contractible into the marked points or onto the boundary. Each arc is considered up to
isotopy. Two arcs are called compatible if they do not intersect in the interior of $\Sigma$.  A maximal collection of distinct pairwise arcs 
is called an {\it ideal triangulation}, see \Fig \ref{fig.disc} for examples. An edge is called external if it is isotopic to a segment of the boundary, otherwise
it is called internal.

It is not hard to get the following formula for the number of total edges
\begin{equation*}
6 g+3b+\#\left| M_{\rm boundary }\right|-6 \ ,
\end{equation*}
where as defined previously $g$ ($b$) is the genus (the number of boundary components) of
$\Sigma$, respectively.
The number of internal edges is $6g+3b-6$, and there are 
a total of $\#\left|M_{\rm boundary}\right|$ external edges.

\subsubsection{The Definition of the Theories}\label{sec.triangulationdef}
Given a triangulation, we have the following rules for defining the 4d $\scN=1$ SCFT:

1. Assign a $SU(N)$ gauge group to each internal edge, and assign a $SU(N)$ flavor group to each external edge.

2. There is a bifundamental chiral superfield between two gauge groups (or flavor groups) if the two corresponding 
edges are in the same triangle of the triangulation. The orientation of the chiral is determined 
by the orientation of the triangle (we take clockwise orientation), i.e, if edge $a$ is before edge
$b$ in the triangle, then the chiral is pointing quiver node $b$.

3. There is a cubic superpotential term for the three chirals in the same triangle.

Although the above definition looks quite different from previous subsection, they actually define the 
same theory. We have summarized the rules in Table \ref{table.rule}.

\begin{table}[htbp]
\caption{Summary of the rule}
\begin{center}
\begin{tabular}{c|c}
\hline
4d gauge theory & triangulation \\
\hline
\hline
$SU(N)$ symmetry & edge \\
\hline 
gauge symmetry & internal edges \\
\hline
global symmetry & external edge \\
\hline
bifundamental and superpotential& triangle \\
\hline
\end{tabular}
\end{center}
\label{table.rule}
\end{table}

\subsubsection{Seiberg Duality is a Flip}\label{sec.SeibergFlip}

\begin{figure}[htbp]
\centering{\includegraphics[width=10cm]{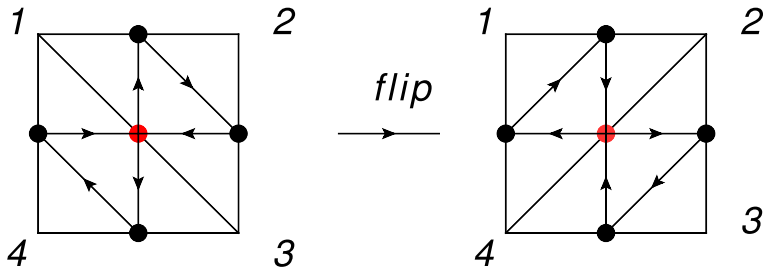}}
\caption{The Seiberg duality is interpreted as flip.}
\label{fig.flip}
\end{figure}

The definition of $\scT_{\Sigma}$ to this point depends explicitly on the choice of the 
triangulation, and we in general have different Lagrangians when we 
choose different triangulations. 
However, different triangulations are related by local moves called flips (\Fig \ref{fig.flip}),
and such a flip is indeed the 
basic Seiberg duality we considered in the last subsection. Therefore, the theories from
different triangulations are related by a sequence of Seiberg duality, 
and our IR SCFT does not depend on the choice of the triangulation.

We can also generalize our definition by replacing the $\scT_{\Delta}$ by $\scT_{\Delta}^l$ theories, 
just as in section \ref{sec.open}.

Finally, let us comment on the relation with
the BPS quivers for the Gaiotto theories (see e.g.~\cite{FST1,Cecotti:2011rv}).
When we dimensionally reduce our
4d $\scN=1$ theory to 1d, we have a quantum mechanics, 
which is almost the same as the 
effective theory of the 1/2 BPS particles of 4d $\scN=2$ SCFT.
In the latter context a Seiberg duality of 4d $\scN=1$ theory is 
interpreted as a quiver mutation which describes the wall crossing phenomena.

There is a major difference,  however.
The difference is that in 4d there is a non-trivial constraint from the existence of 
consistent IR superconformal R-charges.
Indeed, we 
now argue that if the Riemann surface has an empty boundary, 
there are gauge invariant operators with R-charge $0$,
and hence the 4d theory is problematic.\footnote{We would like to thank S.~Razamat for discussion on this issue.}

\begin{figure}[htbp]
\centering{\includegraphics[width=5cm]{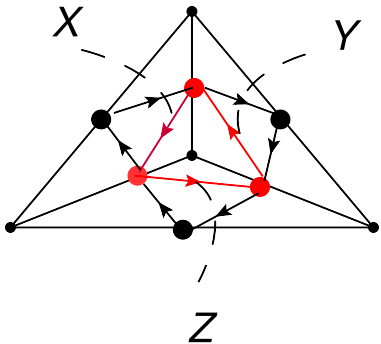}}
\caption{The gauge invariant operator $\Tr(XZY)$ has $r$ charge $0$ in the IR, and violates the unitarity bound.}
\label{fig.gluearound}
\end{figure}

To see this, it is enough to look at a three punctured disc with a bulk puncture. The triangulation and 
the field theory is shown in \Fig \ref{fig.gluearound}.
If require the $\beta$-function to vanish and the superpotential to have R-charge $2$ (more on this in section \ref{sec.R-charges}), 
it is easy to find that the gauge invariant operator $\Tr(XZY)$ has
$R$-charge zero, which invalidates the 4d interpretation. 
However, this theory is perfectly fine 
as the BPS quiver.

\subsection{Bipartite Graph}\label{sec.bipartite}
\begin{figure}[htbp]
\centering{\includegraphics[width=8cm]{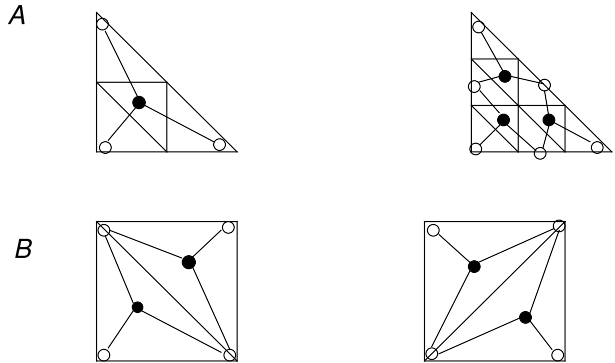}}
\caption{We assign a network to each triangle in the triangulation, and glue them together.}
\label{fig.gluetogether}
\end{figure}

We now show that an ideal triangulation can be translated 
into a bipartite graph defined on the corresponding Riemann surface, 
thus making contact with the contents of section \ref{sec.def2}.
The idea is very simple; we are going to define a network on 
each triangle and define how networks are glued together (\Fig \ref{fig.gluetogether}). 

Let us start with the quiver diagram of the triangle theory.
We place a black (white) vertex at the small triangle 
with clockwise (counterclockwise) orientation. 
We also need to place white vertices on the boundary.
The network is formed by connecting the black and white 
vertices if there is a quiver arrow between them. 
The gluing is achieved by identifying the white vertices. Such constructions
are found in \cite{Xie:2012dw,GoncharovIdeal}. 
We assign a superpotential term to any vertices of the network. 
We can also verify that a flip of the triangulation is equivalent to a square move.

We can compute the permutation for the bipartite graph, following the methods explained in
section \ref{sec.permutations}. For example, if we have a $n$-punctured disc,
then the theory $\scT_{\Sigma}^{l=1}$ gives a permutation $\pi: i\to i+2$ modulo $n$.
We also find that the integer $k$ in \eqref{kdef} is given by $k=2$, and 
hence this permutation  corresponds to a cell of $(\Gr_{2, n})_{\ge 0}$.

\section{Properties of The Theories}\label{sec.properties}

In this section we make preliminary comments on the superconformal IR R-charges
and the  4d superconformal indices of our theories.
More detailed study of our theories will be deferred for
future work.

\subsection{R-charge}\label{sec.R-charges}

Let us next comment on the IR R-charges
for the superconformal $U(1)_R$-symmetry.
It should satisfy the 
following two constraints.

First, the $\beta$-function for Yukawa
     couplings vanish.
This is the same as the requirement that
the  R-charge of the superpotential, and therefore any term in the
     superpotential, is normalized to be 2:
    \beq
\sum_{e\in v} R_e=2 , \ 
\label{sum}
\eeq
for any internal vertex of the quiver diagram.

Second, the $\beta$-functions for the gauge coupling vanish. 
From the NSVZ $\beta$-function, which in our case could be written as
\beq
\frac{d}{d\log \mu}\frac{1}{g_v^2}=
\frac{N}{1-g_v^2 N/8\pi^2}\left[
3-\frac{1}{2}\sum_{e\in v}(1-\gamma_e)
\right] \ ,
\label{gaugebeta}
\eeq
where the anomalous dimension $\gamma_e$ is related to the R-charge
by $\gamma_e=3R_e-2$.
From this condition, we have
\beq
\sum_{e\in V} (1-R_e)=2 \ .
\label{sumdual}
\eeq

In general it is a non-trivial problem to see if there are physical solutions to these equations (recall the discussion
in section \ref{sec.SeibergFlip}).

Let us assume that the bicolored graph is reduced and planar.
We can solve these constraints by assigning parameters 
to zig-zag paths. This is taken from the 
discussion of the torus case in \cite[section 2.4]{Yamazaki:2012cp},
which originally goes back to \cite{Hanany:2005ss}.

Let us assign a $2\pi$-periodic angle $\theta_i$ to each zig-zag path.
We would like to determine the R-charge for a bifundamental field associated $X_e$ with an 
edge $e$ of the bipartite graph.
Now note that
for any edge there are exactly two zig-zag paths going through it.
Then the R-charge for $X_e$ is simply defined to be the relative angles of the two zig-zag paths.
The two conditions \eqref{sum}, \eqref{sumdual}
then follows from the facts that the total angle around a vertex is $2\pi$, and that
the sum of exterior angles of a
polygon is $2\pi$.

Note that this parametrization does not guarantee that the resulting R-charges
are in the physical range. This happens, for example, when the graph is non-reduced;
the R-charge of the bifundamental field is zero when there is a self-intersection of a zig-zag path.

\subsection{Superconformal Index and Spin Chains}\label{sec.index}

Let us also comment on the 4d superconformal index \cite{Romelsberger:2005eg,Kinney:2005ej}
of our theories.

The superconformal index is a twisted partition function (Witten index) on $S^1\times S^3$,
and depends on several chemical potentials.
For 4d $\scN=1$ theories we have two chemical potentials $t, y$ for combinations of spins and energy commuting with the supercharge, and a set of chemical potentials $u_i$ for global symmetries.

Since the index is independent of continuous deformations of the Lagrangian,
UV computation matches with the IR computation. This means this index is 
invariant under the moves of bicolored networks, and is determined purely from the 
IR fixed point.
The invariance of the index under the Seiberg duality and Higgsing
can also be verified explicitly, in exactly the same argument found in \cite{Yamazaki:2012cp},
under the parametrization of section \ref{sec.R-charges}.

In \cite{Yamazaki:2012cp,Terashima:2012cx} a relation between the 4d superconformal index
and the partition function of a 2d integrable spin system has been proposed for the torus case.
It is straightforward to adopt the argument there to the current setup,
and we have a similar relation
\beq
\scI_{\rm 4d}=Z_\textrm{2d spin}  \ . 
\label{I=Z}
\eeq

The spin system on the right hand side of \eqref{I=Z} is defined on a quiver diagram, 
which in our case is realized on the bordered Riemann surface.
The spins reside at the vertices of the quiver diagram, 
and are $U(1)^{N-1}$-valued. These are precisely the 
integral variables of the matrix model representations of the index,
or more physically the Polyakov loop along the thermal $S^1$.

The interactions for the spins are determined by the
1-loop determinants;
the vectormultiplet determinants correspond to the 
self-interaction of the spins, and the hypermultiplet determinants,
associated with an edge of the quiver diagram, represent
the nearest-neighbor interaction among the spins.

The remarkable property of this spin chain in that they
are integrable \cite{Bazhanov:2010kz,Bazhanov:2011mz};
the square move is translated into the double Yang-Baxter move of the 
zig-zag paths (or of strands), which follows from
integrability, i.e. invariance of the partition function under the Yang-Baxter move.
The relation \eqref{I=Z} is reminiscent of the relation
between 4d superconformal index for $\scN=2$ Gaiotto theories
and 2d TQFT \cite{Gadde:2009kb,Gadde:2011ik}.

It would be interesting to work out the reduction of
our 4d superconformal index to 
3d $S^3$ partition function \cite{Dolan:2011rp,Gadde:2011ia,Imamura:2011uw} or 3d superconformal index \cite{Benini:2011nc}.
We expect to find a connection with
the geometry of some hyperbolic 3-manifold,
probably along the lines of \cite{Yamazaki:2012cp,Terashima:2012cx}.

\section{Outlook}

In this paper we defined a new class of 4d $\scN=1$ SCFTs from reduced planar 
bicolored networks. We believe that this has opened up 
an exciting research direction. There are many open problems we could ask on 
our theories, some of which are listed here.

First and foremost, the question is whether or not our theories flow to a nontrivial
IR fixed point.
We have furthermore made several amazing statements from the combinatorial methods, e.g. some 
extremely complicated theories actually flow to the trivial theory. 
It would be interesting
to study the IR properties of our theories by direct physical methods.
For example, can we use the $a$-maximization 
to compute the IR R-charge? What are values of the central 
charges of $a$ and $c$? 
What does the vacuum moduli space look like?
Are there any marginal deformations of the IR fixed point?
What happens if we change the ranks of the gauge group to be unequal,
and cause a cascade of Seiberg dualities?
Do our theories have brane realizations, perhaps with gravity dual?
Can we define similar theories in 3d with Chern-Simons terms?

In addition to the moves preserving the IR fixed point,
we could consider operations on the bipartite graph
which changes the fixed point. An example is to remove an internal edge.
Physically, this turns on an expectation value of the fields represented by this edge. 
Now the theory flows to another fixed point,
and it is remarkably simple to combinatorially identify 
the resulting IR fixed point using the moves and reductions.

For the case of higher genus Riemann surfaces,
we have started with a Riemann surface
and studied interesting networks and the corresponding gauge theories.
However, the story 
will be much more fruitful if we start with a random network on 
Riemann surfaces, to identify the global data charactering a network, 
and to determine if the two theories are the same in the IR. 
We believe our current understanding of the planar network
is a key to the generalization.
Since one may cut the Riemann surface into a disc with boundary 
points, and then gluing back the boundary points in the end,
the understanding of the planar case should 
be really useful in extending the whole story.


In our characterization of the IR fixed point 
we have unexpectedly encountered a cell of the totally non-negative Grassmannian.
However, in our story we have not made use of the actual
geometrical data of the Grassmannian, for example the coordinate on them.
To take this into account we need to incorporate 
weights to the edges of the bipartite graph.
It is tempting to identify this weight with the R-charges with the edge variables.

The same network as in our paper plays an essential role in
a number of physical contexts, including  
the scattering amplitude and 
the wall crossing phenomena. In each case the same mathematics, for example, 
cluster algebras \cite{FZ1,DWZ1}, play crucial roles.
It is natural to ask whether there is any direct physical 
relations between them, i.e. 
whether there is a map between the calculable observables.

\section*{Acknowledgments}

We would like to thank Shlomo S.~Razamat for initial collaboration
and for numerous inputs.
We also thank Nima Arkani-Hamed for stimulating discussion, 
and for sharing some of the results of \cite{ABCGPT}
prior to publication.
This research is supported in part 
by Zurich Financial services membership and 
by the U.S. Department of Energy, grant DE-FG02-90ER40542 (DX),
and by Princeton Center for Theoretical Science (MY).
 

\def\cprime{$'$}
\providecommand{\href}[2]{#2}\begingroup\raggedright\endgroup

\end{document}